\documentclass[a4paper,fleqn]{cas-dc}

\usepackage[super]{natbib}
\usepackage{subcaption}  %  subcaption 
\usepackage{threeparttable} 
\def\tsc#1{\csdef{#1}{\textsc{\lowercase{#1}}\xspace}}
\tsc{WGM}
\tsc{QE}
\tsc{EP}
\tsc{PMS}
\tsc{BEC}
\tsc{DE}

\begin{document}
\let\WriteBookmarks\relax
\def\floatpagepagefraction{1}
\def\textpagefraction{.001}
\shorttitle{Diffusion-based Small Molecule Generation}
\shortauthors{P. Zhang et~al.}

\title [mode = title]{Unraveling the Potential of Diffusion Models in Small Molecule Generation}

\author[1]{Peining Zhang}[type=author,
                        orcid=0009-0003-3094-0220]
% \cormark[1]
\ead{peining.zhang@uconn.edu}
%\address[1]{, Street 129, 1043 NX Amsterdam, The Netherlands}

\author[1]{Daniel Baker}[type=author, orcid=0009-0004-4306-5178]
\ead{daniel.baker@uconn.edu}

\author[2]{Minghu Song}[type=editor, orcid=0000-0003-0887-0767]
\fnmark[1,2]
% \cormark[2]
\ead{minghu.song@ihm.ac.cn}
% \ead{minghu.song@uconn.edu}

\author[1]{Jinbo Bi}[type=editor,
                     orcid=0000-0001-6996-4092]

\cormark[1]
\ead{jinbo.bi@uconn.edu}

\affiliation[1]{organization={Department of Computer Science and Engineering, University of Connecticut},
                addressline={371 Fairfield Road}, 
                city={Storrs},
                postcode={06269}, 
                state={Connecticut},
                country={United States}}
\affiliation[2]{organization={Department of Biomedical Engineering, University of Connecticut},
                addressline={260 Glenbrook Rd}, 
                city={Storrs},
                postcode={06269}, 
                state={Connecticut},
                country={United States}}
\fntext[fn1]{Questions about chemical knowledge can be addressed to M. Song.}
\fntext[fn2]{Minghu Song's current address is 1 Susong Road, Hefei, 230031, Anhui, China }
%\fntext[fn1]{Current affiliation: Institute of Health and Medicine, Hefei Comprehensive National Science Center, Hefei, China.}
\cortext[cor1]{Corresponding author}
% \affiliation[3]{organization={Institute of Health and Medicine, Hefei Comprehensive National Science Center},
%                addressline={1 Susong Road}, 
%                city={Hefei},
%                postcode={230031}, 
%                state={Anhui},
%                country={China}}

% \begin{abstract}
% Diffusion models (DMs) have shown potential utility in drug discovery by facilitating molecule generation. This paper presents a comprehensive review of DMs for small molecule generation, categorizing them into a taxonomy based on both their mathematical foundations and chemical applications. We investigate models that handle 1D, 2D, and 3D molecular representations with and without a protein binding target, and analyze current challenges focusing on those encountered in generating 3D valid structures. Particularly, extensive experiments are performed on benchmark datasets using a unified benchmarking approach to compare the generation performance of existing 3D methods. We conclude that current DMs seldom generate valid molecules and provide insights into future directions for improvement.
% \end{abstract}
% New abstract： 
\begin{abstract}
 % Generative AI offers chemists new ideas for drug design and aids in exploring vast chemical spaces. Diffusion models (DMs), an emerging tool, have gained much attention in drug R&D recently. This paper comprehensively reviews the latest advances and applications of DMs in molecular generation. It begins by introducing the theoretical foundations of DMs, then classifies various DM molecular generation methods based on math and chemical applications. The review further explores the performance of these models on benchmark datasets, especially compares the generation performance of existing 3D methods, and concludes by highlighting current challenges and proposing future research directions to fully unleash DMs' potential in drug discovery.
 Generative AI presents chemists with novel ideas for drug design and facilitates the exploration of vast chemical spaces. Diffusion models (DMs), an emerging tool, have recently attracted great attention in drug R\&D. This paper comprehensively reviews the latest advancements and applications of DMs in molecular generation. It begins by introducing the theoretical principles of DMs. Subsequently, it categorizes various DM-based molecular generation methods according to their mathematical and chemical applications. The review further examines the performance of these models on benchmark datasets, with a particular focus on  comparing the generation performance of existing 3D methods. Finally, it concludes by emphasizing current challenges and suggesting future research directions to fully exploit the potential of DMs in drug discovery.
\end{abstract}

% \includegraphics{cas-grabs.pdf}

% •
% AlphaFold models have rigid protein structures.
% •
% Advanced deep algorithms fed relevant data can accurately predict structures.
% •
% Algorithmic advances can harness AlphaFold in drug discovery.
% •
% Amplifying the active state can broaden and enrich the input to virtual docking.
% •
% The ‘reverse pharmacophore’ of ligand-centered receptors can minimize toxic side effects.
% •
% Drug discovery projects can be benefited from the machine learning and deep learning based scoring functions.
% •
% The current review highlights various success stories in the identification of lead compounds using such scoring functions which are verified from experimental bioassay and characterization studies.
% •
% Importance of accurate protein structure prediction for small molecule drug design.
% •
% Key structural features and properties used in small molecule drug design.
% •
% Computational tools and methods for predicting and optimizing small molecule structures.
% •
% Examples of successful small molecule drug design based on molecular structure.

% \begin{highlights}
% \item Diffusion models show promise in small molecule drug design.
% \item Diffusion models are generative models applied across various molecular representation approaches.
% \item Unified benchmarking exposes limitations in current diffusion model approaches for molecular design
% \end{highlights}

\maketitle

\section{Introduction}

In recent years, deep generative models have emerged as a new frontier in {\em de novo} molecular design. Among these, diffusion models (DMs), inspired by non-equilibrium statistical physics,\cite{dhariwal2021diffusion,ho2020denoising} have shown remarkable potential in modeling the three-dimensional geometric structure of molecules. In the field of image generation, these emerging models have surpassed earlier generative models such as variational autoencoders (VAEs) and generative adversarial networks (GANs) in terms of both the fidelity of the generated images and training stability. This success has encouraged researchers to further explore their applicability in the structural generation of small molecules. 

This investigation holds notable importance as the conventional drug discovery process encounters a multitude of challenges, such as extended timelines, substantial financial demands, and elevated rates of failure. Artificial intelligence, particularly through generative models, offers a promising avenue to enhance efficiency across various phases of this intricate procedure. Virtual screening, for instance, is a critical component that can occupy up to half of the drug research and development cycle, representing a considerable bottleneck.\cite{stokes2020deep} The utilization of diffusion models to generate diverse candidate libraries of candidate molecules for virtual screening is therefore a rapidly advancing and promising technique, with the potential to improve both the effectiveness and efficiency of this crucial workflow.

\subsection{Molecule Representations}
\begin{figure*}
	\centering
	\includegraphics[width=1.0\textwidth]{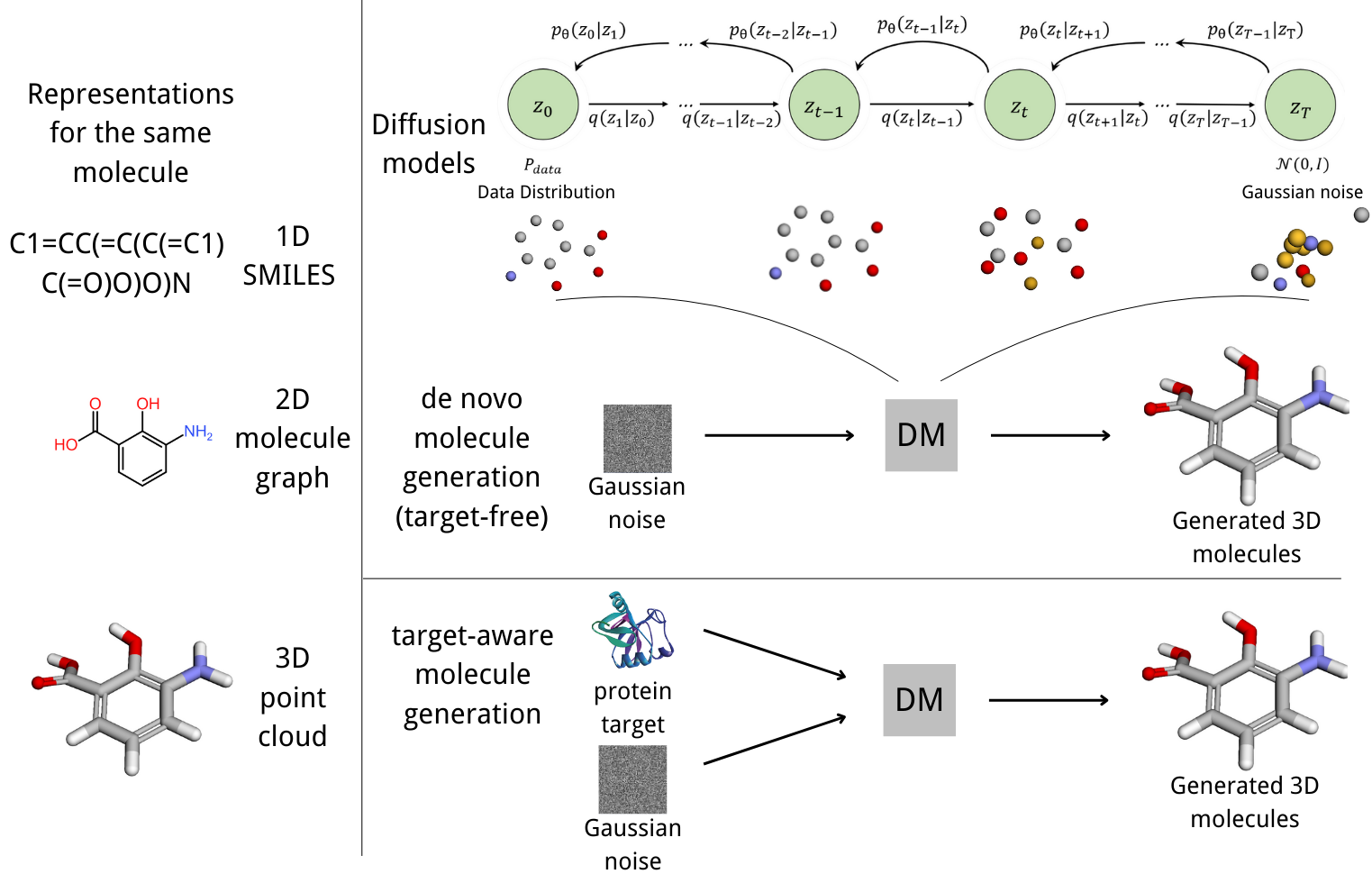}
	% \caption{Left: molecular representations; Right: the diffusion model generative process (top), and how DM generates conformation from a given molecular graph (middle) and generates a new 3D molecule (bottom). }
    \caption{Left: 1D/2D/3D molecular representations; Right: the diffusion model generative process, consists of steps of forward diffusion process (from left to right) and backward diffusion process (from right to left), described by Gaussian process (top), and how DM generates a new 3D molecule (middle) and  generates 3D molecule based on protein target (bottom). }
	\label{fig:mol_repre}
\end{figure*}
% \refstepcounter{figure}
% \label{fig:mol_repre}
 Molecules can be encoded and generated in a variety of ways, as shown in Figure \ref{fig:mol_repre}. The Simplified Molecular Input Line Entry System (SMILES)\cite{weininger1988smiles} represents atoms and their chemical bonding environments using textual symbols, allowing for molecular generation with Natural Language Processing (NLP) techniques.\cite{edwards2022translation} Two-dimensional molecular graphs, in which nodes symbolize atoms and edges represent chemical bonds, are also common representations. However, neither of these methods can fully capture the three-dimensional information of molecules. In contrast, in computer science, point clouds or voxels can better represent the three-dimensional connectivity and spatial arrangement (atomic positions) of molecules. Additionally, many models for molecular generation have considered rotational and translational equivariance to better handle three-dimensional molecular representation.\cite{fuchs2020se, satorras2021n}

DMs have become a powerful tool for molecular geometry generation.\cite{xu2022geodiff, hoogeboom2022equivariant} They operate via a two-step process.\cite{ho2020denoising, 51347} In the forward diffusion step, noise is gradually introduced to an initial, clean molecular representation, a molecular graph, or a set of coordinates of the most stable conformation. As more noise is incrementally added, the molecule's representation becomes increasingly perturbed. The reverse diffusion step, on the other hand, aims to remove this noise gradually. The model can generate new molecular geometries by learning the noise addition and removal patterns. It does this by iteratively predicting the previous, less noisy state at each step of the reverse process until a realistic molecular structure is obtained. Unlike traditional molecular geometry generation methods that rely on rule-based systems or sampling from pre-defined libraries, DMs can incorporate large amounts of data during training. This enables them to capture the complex, non-linear relationships within molecular structures. As a result, DMs can generate novel molecular geometries that are difficult to access through traditional means. They are employed for conformation generation tasks, such as predicting the three-dimensional arrangement of atoms for a given molecular graph or SMILES,\cite{xu2022geodiff} and directly generating molecules with their corresponding conformations or specific geometric properties. For example, DMs can create molecules that are likely to exhibit high binding affinity to a target protein.\cite{guan20233d_targetdiff,huang2024dual}

This review aims to comprehensively analyze the application of various emerging DMs in small-molecule generation. First, it begins by elucidating the fundamental mathematical principles, including the crucial forward and reverse diffusion processes, examining the noise addition and removal mechanisms, with a focus on how these processes are adapted for molecular generation and incorporate equivariance to preserve essential geometric properties. Next, a detailed taxonomy (Figure \ref{fig:taxonomy}) of these models is created, classifying them based on the presence of protein-binding targets, molecular representations (1D, 2D, 3D), and diffusion frameworks.  This taxonomy provides a structured framework to help researchers navigate the extensive literature and select appropriate methods for their specific molecular generation tasks. Finally, the performance of these models across various applications is comprehensively evaluated using benchmark datasets. The challenges in small-molecule generation are discussed in detail and precisely identified to inspire further research. Future directions are proposed to unlock these models' potential in drug discovery fully.

\section{The Proposed Taxonomy}

We propose a new taxonomy to categorize the DMs for molecular generation up to August 2024. Figure \ref{fig:taxonomy} organizes the different DM approaches into the `target-free' and `target-aware' categories first. Target-free models are primarily utilized for exploring broad chemical space and generating valid and diverse molecular structures without prior knowledge of a specific biological target. In the context of this taxonomy, they are often considered foundational; they serve to validate the core capabilities and technical feasibility of new DM methodologies and can underpin the development of various downstream, more specialized conditional generation models. In contrast, target-aware models are designed to utilize specific information, such as the 3D structural details of a biological target. This allows them to generate molecules with a higher potential for efficacy and specificity towards the intended target like protein, which is a significant advantage in drug design. Following this primary distinction, the taxonomy then distinguishes these approaches further based on whether they focus on the generation of the conformation or engage in {\em de novo} molecule generation.Conformation generation starts with an existing molecular graph and refines its atomic spatial arrangement, whereas {\em de novo} generation creates entirely new molecular structures from scratch. 

The third level of categorization focuses on data representation, which determines whether the molecules are generated in a 1D, 2D, or 3D representation.  %3D structural methods require models capable of precisely positioning atoms in space, respecting physical constraints. 2D graph-based approaches represent molecules as nodes (atoms) and edges (bonds), typically employing graph neural networks to capture atomic relationships. 1D SMILES-based methods use string representations, offering efficiency in molecular structure generation and processing through sequence models like transformers.
DM methods are further categorized according to the specific formulation of the DM process.  Two main DM formulations include the denoising diffusion probabilistic model (DDPM) and the score-based DM (SDM), which defines and estimates a score related to the data log-likelihood. Because 3D molecular structures remain unchanged under coordinate translation and rotation, both DM formulations have modifications that allow the molecular generation to be equivariant with respect to SE(3), the Special Euclidean group in three dimensions, which consists of simultaneous rotations and translations in three-dimensional space. For instance, methods based on equivariant graph neural networks (EGNNs) can ensure rotation and translation equivariance, while some others may only provide permutation equivariance. Thus, the fifth level categorizes the methods based on their equivariance properties and network architectures.

\begin{table*}[h!]
\centering
\begin{tabular}{|p{1.25cm}|p{2cm}|p{1.25cm}|p{1.25cm}|p{2cm}|p{6.25cm}|}

\hline
\textbf{Target} \newline \textbf{Specificity} & \textbf{Generation} \newline \textbf{Config.} & \textbf{Molecule} \newline \textbf{modality} & \textbf{Diffusion Formulation} & \textbf{Symmetry Consideration and Network Architecture} & \textbf{Model Name} \\ \hline
{\textbf{Target free}} 
 & Conformation \newline generation & 3D point cloud& Score-based & SE(3)-eq\newline GNN & GeoDiff\cite{xu2022geodiff}, Torsional Diffusion\cite{jing2022torsional} \\ 
 \cline{5-6} 
& & & & non-SE(3)\newline GNN & MCF\cite{wang2024swallowing}\\ 
\cline{2-6}

 & Molecule\newline generation & \multirow{2}{1cm}{3D point cloud/ voxel}& DDPM & SE(3)-eq\newline GNN & EDM\cite{hoogeboom2022equivariant}, HierDiff\cite{qiang2023coarse},MUDiff\cite{hua2024mudiff}, GeoLDM\cite{xu2023geometric}, EQGAT-diff\cite{lenavigating}\\  \cline{5-6} % 

  & & & & non-SE(3)\newline CNN & VoxMol\cite{o20243d_voxmol}, FMG\cite{dumitrescu2024field}\\ \cline{4-6}
 
 & & & Score-based & SE(3)-eq\newline Graph transformer & MiDi\cite{vignac2023midi}, MolDiff\cite{peng2023moldiff}, GFMDiff\cite{xu2024geometric}, JODO\cite{huang2023learning}\\  \cline{5-6}
 
 & & & & SE(3)-eq\newline GNN & MDM\cite{huang2023mdm}\\
 \cline{3-6}
 &  & 2D graph& Score-based & Perm-eq\newline Graph transformer & DiGress\cite{vignac2022digress}, FreeGress\cite{ninniri2024classifier}, GLDM\cite{wang2024gldm} \\ \cline{3-6} 
 &  & 1D SMILES& DDPM & -\newline Text transformer & TGM-DLM\cite{gong2024text}, DIFFUMOL\cite{peng2024hitting} \\ \hline

 {\textbf{Target aware}}& Conformation generation (Protein-ligand docking)& 3D point cloud & Score-based & non-SE(3)\newline GNN & DiffDock\cite{corso2022diffdock}, Re-Dock\cite{huang2024re}, PocketCFDM\cite{voitsitskyi2023boosting} \\ \cline{2-6}
& Molecule  generation (Ligand generation) & 3D & DDPM & SE(3)-eq\newline GNN & TargetDiff\cite{guan20233d_targetdiff}, DiffSBDD\cite{schneuing2022structure}, DiffLinker\cite{igashov2024equivariant}, DecompDiff\cite{guan2024decompdiff}, SILVR\cite{runcie2023silvr}, PMDM\cite{huang2024dual}, MolSnapper\cite{ziv2024molsnapper}, LinkerNet\cite{guan2024linkernet}, BindDM\cite{huang2024binding}, IPDiff\cite{huang2023protein}, PromptDiff\cite{yang2023prompt}, D3FG\cite{lin2024functional}, AutoDiff\cite{li2024autodiff}, ShapeMol\cite{chen2023shape}, STRIDE\cite{zaman2023stride}, InterDiff\cite{wu2024guided}, KGDiff\cite{qian2024kgdiff}, DiffDec\cite{xie2024diffdec}, AliDiff\cite{gu2024aligning} \\ \cline{5-6} 
 &  &  & & non-SE(3)\newline GNN & AutoFragDiff\cite{ghorbani2023autoregressive} \\ \hline
\end{tabular}
\caption{A comprehensive taxonomy of diffusion models for molecular generation, categorizing existing approaches based
on target awareness, generation type, data representation, model equivariance characteristic, model architecture.}
\label{tab:taxonomy2}
\end{table*}

% \refstepcounter{figure}
% \label{fig:taxonomy}

\section{Overview of Diffusion Models}
%introduce DMs from theory forward and reverse to training to sampling.
%DMs are generative models that iteratively recover a data distribution starting from a simple, often Gaussian, noise distribution. 
DM applies `diffusion' to the training data where `diffusion' refers to the complete process of gradually adding noise to data (forward diffusion) and then learning to reverse that process to recover the original data by progressively removing the noise (reverse diffusion). The reverse diffusion is implemented by a neural network which can predict the noise of each step thus iteratively recover the data distribution from a commonly Gaussian noise distribution by interpolating between a noise distribution and the desired data distribution $p(\mathbf{x})$.\cite{yim2024diffusion} There are two main DM formulations: DDPM with a  discrete time scheduling, and SDM derived from a continuous-time formulation.
\subsection{Basic Formulations} As shown in the top-right part of Figure
\ref{fig:mol_repre}, DDPM defines a theoretical forward diffusion $q(\mathbf{x}_t | \mathbf{x}_{t-1})$ that sequentially adds noise to data, transforming the original data $\mathbf{x}_0$ into standard Gaussian noise $\mathbf{x}_T$ through T discrete timesteps. Theoretically, the reverse diffusion process $q(\mathbf{x}_{t-1} | \mathbf{x}_t,\mathbf{x}_0)$ is also a Gaussian distribution used in derivation, which is estimated by a neural network in DDPM.  The network learns a distribution  $p_\theta(\mathbf{x}_{t-1} | \mathbf{x}_t) = \mathcal{N}(\mathbf{x}_{t-1}; \mu_\theta(\mathbf{x}_t, t), \Sigma_\theta(\mathbf{x}_t, t))$, which depends only on $\mathbf{x}_t$ rather than both $\mathbf{x}_t$ and $\mathbf{x}_0$, making the estimated reverse process Markovian.
During training, the model learns to estimate the noise by minimizing the Kullback-Leibler divergence between the theoretical and estimated reverse processes. This optimization can be simplified to minimize the mean squared error between the theoretical and estimated means of the two Gaussian distributions. When generating new samples, $T$ reversed steps are performed.

SDM and its various extensions\cite{song2019generative, 51347} are derived from continuous-time formulations, where both forward and reverse processes are defined by Stochastic Differential Equations (SDEs). The key innovation is to learn a score function defined as $s_\theta(\mathbf{x}) = \nabla_{\mathbf{x}} \log p_{data}(\mathbf{x})$, the gradient of the log-likelihood of the true data distribution. This score function is learned through a neural network via score matching\cite{hyvarinen2005estimation} and captures how the probability density of real data samples changes in the data space. Instead of directly modeling the complex probability distribution, this approach estimates its gradient, which is often simpler and more efficient to learn. During sampling, the learned score function guides the generation process, allowing differential equation solvers (e.g., the Euler method) to trace the sampling trajectory and recover the final sample $\hat{\mathbf{x}}_0$.

These DMs offer several advantages for molecule generation. Their well-defined mathematical framework enables precise analysis of key properties such as likelihood bounds, sample quality, and generation diversity. Their flexibility allows adaptation to various data types, particularly valuable in drug discovery where molecular representations span continuous (atom coordinates) and discrete (atom types) domains. Unlike autoregressive NLP models (commonly used to generate SMILES), DMs generate the entire molecular structure simultaneously, simplifying the generation of complex molecular structures.

\subsection{Equivariant Diffusion Models}
%What is equivariance -> Why equivariance -> DM have equvairance
%Equivariance refers to a function $f$ transforms output consistent with input transformations, such that $f(T(x)) = T(f(x))$. In contrast, an invariant function satisfies $f(T(x)) = f(x)$. When $f$ outputs a scalar, invariance implies equivariance, making it a natural approach to achieving equivariance.  For molecular modeling, a molecule's representation should remain consistent under different coordinate positions and orientations in 3D space. 

A function $f$ is equivariant with respect to a transform $T$ if and only if $f(T(x)) = T(f(x))$. In contrast, an invariant function satisfies $f(T(x)) = f(x)$. When $f$ outputs a scalar, invariance implies equivariance, so ensuring invariance will achieve equivariance. For molecular modeling, a molecule's property should remain consistent under different coordinate positions and orientations in 3D space. 

A set of DMs achieve SE(3)-equivariance when generating molecules, meaning they are equivariant with respect to coordinate rotation and translation (SE(3) transformations). With equivariance, models better capture the physical properties of molecules.\cite{fuchs2020se, satorras2021n} To enforce equivariance, a key technique is to measure Euclidean distances of atomic coordinates in a neural network instead of exact coordinates, because Euclidean distance is invariant under rotation and translation.

The advantages of equivariance are significant. By enforcing physical symmetries, these models reduce the search space of possible solutions and allow more efficient exploration of the chemical space.\cite{fuchs2020se} They provide more reliable predictions across different molecular orientations, which is crucial for applications like molecular docking, where consistent predictions across molecular conformations are essential.

\subsection{Diffusion Models for Molecule Generation} \label{sec:DMMG}
The foundation of applying DMs to molecule generation involves encoding both the continuous atom coordinates and the categorical features such as atom and bond types. %For each a a molecule, both the type and coordinates of each atom need to be determined. 
The atom coordinates, representing by a matrix $X = [x_1, x_2, ... , x_n]^T$ $\in$ $\mathbb{R}^{n \times 3}$, can be computed by a direct diffusion process.  For the categorical atom types (and bond types), a common strategy enabling the use of a DM is to convert the one-hot encoding to a probabilistic encoding that is a function of latent embeddings of atom types and the diffusion is applied to the continuous latent embedding.\cite{hoogeboom2022equivariant} Each element of the probabilistic embedding vector, $p(h|z^{(h)}_0)$, corresponds to an atom type $h \in \{h_1,...,h_d\}$, indicating the likelihood of taking the type $h$, where $z^{(h)}_0$ is the recovered embedding at $t=0$ from the diffusion process. Typically, a Gaussian distribution \(\mathcal{N}(u | z_0^{(h)}, \sigma_0)\)  with mean \(z_0^{(h)}\) and a pre-specified  variance \(\sigma_0^2\) is used and $p(h|z^{(h)}_0)$ can be computed as follows: %For continuous features such as atom positions, a common approach is representing them as atom coordinates(i.e. conformations) matrix  The coordinates matrix is computed by the denoising. For discrete features like atom types, A common encoding approach applied by the original equivariant DM called EDM\cite{hoogeboom2022equivariant} involves estimating category probabilities through integrals over Gaussian distributions. For categories $\{c_1,...,c_d\}$ representing atom types, the probability of category $h$ can be defined as probability parameters $\mathbf{p}$, which is proportional to the integral:
\begin{equation}
p(h|z^{(h)}_0) \propto	 \int_{1-\frac{1}{2}}^{1+\frac{1}{2}} \mathcal{N}(u | z_0^{(h)}, \sigma_0) du \label{eq:zeroth likelihood} 
\end{equation}
where the integration interval $[1-\frac{1}{2}, 1+\frac{1}{2}]$ is chosen because, in one-hot encoding, the active class has a value of 1. %The model employs the diffusion process on a latent variable $z_t^{(h)}$. The denoising result $z_0^{(h)}$ serves as the mean of the Gaussian distribution.

\subsection{Neural Network Architectures}
% {\bf potentially mention 3D molecular graph and define edges correspond to bonds[Update]:introduce edge and 3D molecular graph}
The reverse process of DM aims to recognize and gradually remove noise from data at each timestep, with input and output maintaining the same dimension and shape. Initially, DDPM\cite{ho2020denoising} used the U-Net\cite{ronneberger2015u} architecture, which was later supplanted by transformers for improved performance and scalability\cite{peebles2023scalable}.

Graph Neural Networks (GNNs) are particularly suited for molecular modeling as they naturally capture atomic relationships and molecular structure.\cite{zhou2020graph} Message Passing Neural Networks (MPNNs), a foundational type of GNN, introduced permutation-equivariant graph processing through message passing and aggregation that operate consistently across nodes and their neighbors.

Equivariant Graph Neural Networks (EGNNs)\cite{satorras2021n} extended this MPNN approach by implementing rotation and translation equivariance for 3D representations. In EGNNs, a 3D molecular graph is constructed based on  interatomic distances, where edges connect atoms, often corresponding to chemical bonds. This architecture, first used and implemented in equivariant DM (EDM),\cite{hoogeboom2022equivariant} has become dominant in molecular generation. Graph transformers, inspired by advances in NLP, use the multi-head attention mechanism to model complex dependencies between atoms, representing different interactions. Recent models such as MUDiff\cite{hua2024mudiff} and InterDiff\cite{wu2024guided} applied equivariant graph transformers to 3D molecular generation.

\section{DM-based {\em de novo} Molecule Generation}
\subsection{Target-free Generation}
Target-free generation models are typically trained and evaluated on benchmark datasets containing 3D molecular structures. Among these, the QM9 dataset\cite{ramakrishnan2014quantum} is widely used, providing computed geometric, energetic, electronic, and thermodynamic properties. It supports both unconditional generation and the generation of molecules that optimize certain molecular properties, such as geometric configurations or electronic characteristics. Additionally, it ensures molecular diversity and novelty, measured by the difference between generated structures and those in the training set. GEOM-DRUGS dataset\cite{axelrod2022geom} is another dataset used to create models for generating larger, drug-like molecules. On the other hand, the GuacaMol dataset\cite{brown2019guacamol} is often applied to 2D graph-based models, containing 1.3 million drug-like molecule graphs.
\paragraph{Molecule Conformation Generation}
The conformation generation task aims to estimate atom coordinates from a given molecular graph, where DMs are first applied to molecular generation tasks. GeoDiff\cite{xu2022geodiff} is the first work that employs SE(3)-EGNN to model the spatial stability. It computes a matrix of atom coordinates by denoising from randomly sampled coordinates from a Gaussian distribution. However, GeoDiff suffers from computational inefficiency, requiring 1000 inference steps, making it impractical for real-time applications or large-scale molecular screening. Torsional Diffusion\cite{jing2022torsional} enhances GeoDiff in terms of both effectiveness and efficiency by shifting the focus from atom positions $C \in \mathbb{R}^{n \times 3}$ to torsion angles $T \in [0,2\pi)^m = \mathbb{R}^{m}$, where $m$ is the number of flexible torsion angles. As a result, the search space is significantly reduced. It allows inference to start from existing conformations generated by other tools, such as RDKit,\cite{rdkit} requiring only 10-20 inference steps compared to GeoDiff's 1000 steps. Despite this improvement, Torsional Diffusion relies on rule-based methods to identify rotatable bonds, which may fail for complex molecules. The direct joint diffusion on torsion angles can also lead to steric clashes and accumulated coordinate displacement, complicating the denoising process. Molecular Conformer Fields (MCF)\cite{wang2024swallowing} challenges the requirement for equivarianceby directly training a graph-transformer network on atomic coordinates without enforcing strict equivariance constraints. Instead, MCF incorporates equivariant data augmentation to encourage equivariance implicitly. This approach emphasizes that while equivariance introduces beneficial inductive biases, model scalability remains equally important. This approach demonstrates that increased model capacity can significantly enhance generalization even without strict enforcement of rotational equivariance.

\paragraph{Molecular Structure Generation}
% Unconditional methods often rely on extensive chemical libraries like QM9\cite{ramakrishnan2014quantum} and GEOM-drugs\cite{axelrod2022geom} that provide a broad range of molecular structures. 
Molecular Structure Generation aims to directly generate valid, stable, and novel molecules in their appropriate conformations without relying on a given molecular graph.
EDM\cite{hoogeboom2022equivariant} adopts GeoDiff's representation of (continuous) atom coordinates, but further pioneered the generation of discrete molecular features as summarized in Sec. \ref{sec:DMMG}. 
%It estimates the probability of each atom type category by integrating the Gaussian distribution {\bf I do not understand the previous sentence}, effectively handling the discrete nature of molecular structures within the continuous framework of DMs. 
However, EDM relies on external software (e.g., OpenBabel\cite{o2011open}) to determine bond types based on the generated atom types and coordinates by the DM. While effective, it leaves room for more integrated modeling strategies.  %Published shortly after EDM, {\bf what does MDM stands for? [Updated]}Molecular Diffusion Models (
MDM\cite{huang2023mdm} tackles this issue by enhancing the EDM model architecture. It integrates additional encoders: one is dedicated to encoding local edges capturing covalent bonds and other short-range interactions, while the other is designed for modeling global edges, accounting for longer-range forces (e.g., van der Waals interactions). Modeling the longer-range interactions benefits the generation of larger molecules. However, MDM's dual-encoder architecture struggles with scalability issues when generating very large molecules due to the quadratic scaling of pairwise interactions.

\paragraph{3D  Molecular Graphs}
Expanding from EDM and MDM, several works have aimed to build a more unified model by explicitly generating bonds alongside atom coordinates, and atom types, which effectively constructs complete molecular graphs. A critical limitation shared by all explicit bond modeling approaches, is the computational burden of maintaining N×N bond predictions. While transformer-like architectures can derive this information from attention matrices with reasonable computational cost, direct generation of bond matrices scales as O($N^4$), making it impractical for large molecules. 
%Since these 3D molecule generation methods represent 3D structures as graphs and use EGNN\cite{satorras2021n} to encode and generate graphs of 3D structures.
MiDi,\cite{vignac2023midi} building upon EDM, presents an end-to-end approach for generating 3D molecular graphs. MiDi directly predicts bond alongside atom types and coordinates through a single diffusion process. To address the interdependency between these features, MiDi assigns different diffusion rates that prioritize predicting atom coordinates before bond types and atom types. The main weakness of MiDi lies in its heuristic approach to setting different diffusion rates for different molecular features. This manual tuning lacks theoretical justification and may not generalize well across different molecular types or sizes.  Like MiDi, MUDiff\cite{hua2024mudiff} extends EDM to enable simultaneous generation of both molecular graphs and conformers. MUDiff introduces MUformer, a novel Transformer that ensures invariance to SE(3) transformations as well as reflection, which is called E(3)-equivariant.  MUformer jointly learns 2D and 3D molecular structures by encoding atomic, positional, and structural information. To capture atom relations such as connectivity and spatial arrangement, MUformer  encodes two matrices of $N\times N$, respectively, reflecting 2D binary connectivity in the molecular graph and 3D spatial proximity, among $N$ atoms. %This matrix encodes pairwise relationships between atoms, reflecting both their bonding patterns and spatial distances.
EQGAT-diff\cite{lenavigating} systematically explores the design space of SE(3)-equivariant diffusion models, introducing novel network architectures to represent discrete bond types and evaluating the interplay between continuous and discrete state spaces. The work also addresses important design considerations including $\mathbf{\epsilon}$- vs. $\mathbf{x}$-parameterization, time weighting strategies, and transferability issues.
Similarly, JODO\cite{huang2023learning} introduces the Diffusion Graph Transformer (DGT), which shares similarities with MUformer in jointly modeling 2D molecular graphs and 3D structures. %DGT uses node, edge, and position inputs, applying dot-product attention to update these features, thereby encoding both graph connectivity and spatial geometry. {\bf the above sentences feel very hand wavy without the real content. In what exact way JODO models both 2D and 3D info?[Updated]} 
However, JODO distinguishes itself in the sampling step by employing a self-conditioning mechanism,\cite{chen2022analog} which leverages the connectivity matrix of the predicted final result estimated from previous sampling steps to guide subsequent generation. This mechanism ensures consistency between the generation directions, improving the fidelity of the generated graph structures. But this mechanism also introduces potential error accumulation, where incorrect predictions in early sampling steps can propagate and compound throughout the generation process.

\paragraph{Multi-stage Generation}
Multi-stage approaches use multiple sampling processes to generate molecules rather than a single process, which usually provides better scalability for larger and more complex molecules. It is important to note that the definition of "stages" varies across works, and the classification is primarily based on the presence of distinct generation phases. For example, HierDiff\cite{qiang2023coarse} employs a hierarchical generation process. In its first stage, a coarse-grained generator within DMs assembles molecular fragments to a coarse molecule graph. Then, a fine-grained iterative refinement generator refines these fragments into atom-level structures. 
%, ensuring high molecule stability. {\bf does this mean noise was first added to fragment embedding, and then the fragment embedding is used to generate atom-level graph? [Answer]: Denosing process generate fragment embeddings, and then the fragment embedding is used to generate atom-level graph} 
MolDiff\cite{peng2023moldiff} fully integrates its two-stage approach for fragments and atoms 
%{\bf here the two stage does not mean fragment vs. atom stages, right? this means atom vs. bond stages? [Answer]: Yes} 
within the DM framework. Its first stage performs a similar calculation as EDM to generate atom types and positions but replaces EDM's random atom type with a sample from a dummy type `none-type` as the prior distribution.
% {\bf this is not clear. Do not know what you want to say. [Updated]}
The second stage generates bond types and refines atom types and positions. 
% {\bf is this theoretical foundation different from Sec 2? [Updated]}{\bf I really have no idea about what kind of control and adjustment do you mean here. [Updated]} 
GeoLDM,\cite{xu2023geometric} inspired by LDM(Latent Diffusion Model) for image generation,\cite{rombach2022high} presents a path toward a unified model capable of generating molecules of varying sizes, similar to how DMs for images can handle different image sizes and complexities. 
% {\bf I never understand what kind of flexibility of DM does it mean? [Updated]} 
GeoLDM consists of two key components: a variational autoencoder\cite{kingma2013auto} (VAE) for molecule encoding-decoding, and a DM trained on the VAE's latent space. During sampling, generated latent vectors are mapped back to molecular structures through the VAE decoder. This combination simplifies the DM module and facilitates future improvements. GFMDiff\cite{xu2024geometric} extends latent DM by generating atoms and bonds directly. It introduces a Dual-Track Transformer Network (DTN) with separate tracks for modeling atomic and bond features, enabling accurate prediction of atom types, positions, and bond valencies.
A fundamental limitation of multi-stage approaches is their inherent susceptibility to error accumulation across different generation phases. Unlike image generation where VAEs\cite{rombach2022high} can achieve high-fidelity reconstruction while significantly compressing information, molecular generation methods are constrained by SE(3) equivariance requirements that prevent such effective compression. This constraint forces each stage to operate with potentially lossy representations, leading to inevitable error propagation between stages that can compromise the final molecular structure quality.

\paragraph{Voxel-based Generation}
VoxMol\cite{o20243d_voxmol} explores a voxel-based representation for molecules, modeling atoms as continuous densities within a 3D grid. In this representation, the spatial coordinates of atoms are implicitly encoded by the grid structure, while atomic properties (such as atom types) are represented as continuous density values. For the diffusion process, VoxMol employs a 3D U-Net architecture, a variant of the U-Net\cite{ronneberger2015u} commonly used in image segmentation tasks. This model enables efficient sampling while maintaining competitive performance compared to EDM.\cite{hoogeboom2022equivariant} FMG\cite{dumitrescu2024field} extends VoxMol's voxel representation to continuous tensor fields, modeling atomic properties like electron densities as smooth tensor fields over a 3D grid. By applying sample-specific field orientation and leveraging U-Net's translation equivariance, the model achieves SE(3)-equivariance, allowing more accurate capture of molecular structures.
The voxel-based paradigm faces fundamental discretization challenges, where representing continuous molecular structures on discrete grids introduces approximation errors and creates a trade-off between computational efficiency and spatial accuracy. Additionally, these approaches have received limited research attention compared to point cloud-based methods, making their long-term viability and mainstream adoption potential uncertain.

\paragraph{2D Graph-based Models}
Graph-based models differ fundamentally from 3D molecular graph approaches by using discrete diffusion processes rather than diffusion in atom coordinate spaces. Before the adoption  of DMs, various strategies have been explored, including simultaneous generation of atoms and adjacency matrices,\cite{simonovsky2018graphvae} sequential atom-by-atom decoding,\cite{de2018molgan} and substructure-based generation.\cite{jin2018junction} DiGress\cite{vignac2022digress} introduces a discrete denoising diffusion approach for graph generation, where the diffusion process is represented by transition matrices that control edge addition/removal and node/edge category modifications. A graph transformer network is then trained to reverse this process, transforming intermediate states into a series of node and edge classification results that correspond to the transitions, ultimately generating the final graph. FreeGress\cite{ninniri2024classifier} extends DiGress by eliminating the need for an external classifier during generation,\cite{dhariwal2021diffusion, ho2022classifier} simplifying the generation process. GLDM\cite{wang2024gldm} introduces a Graph Latent DM with an encoder-decoder architecture. It encodes molecular graphs into lower-dimensional latent representations, applies DMs to this latent representation, and then the generated embedding are used to decode back to molecular structures. Operating in the latent space simplifies the diffusion process and improves training efficiency.

\paragraph{1D SMILES-based Models}
Generating molecules based on SMILES relies on NLP techniques after tokenizing the chemical symbols. DIFFUMOL\cite{peng2024hitting} introduces a DM that operates in an embedding space, with a tokenizer decoding the generated embedding vectors into SMILES strings. This approach is similar to DiffusionLM\cite{li2022diffusion} for general text generation. DIFFUMOL's training objective minimizes two losses: a denoising loss $\|z_0 - f_\theta(z_t, t)^2\|$  for the diffusion process, and a rounding loss $\|Embedding(y) -f_\theta(z_1, 1)\|$ for tokenization. TGM-DLM\cite{gong2024text} proposes a text-guided molecule generation model that converts natural language descriptions of molecules into SMILES strings, following MolT5's\cite{edwards2022translation} task setup of translating between natural language and SMILES. It combines a translation module with a `Correction` module (a separate DM) to refine SMILES grammar, using  both a network architecture and a DM-based generation approach similar to those in DIFFUMOL.
A fundamental limitation shared by both 1D SMILES-based and 2D graph-based models is their reliance on datasets that differ significantly from those used by 3D-based approaches, making direct performance comparisons challenging and hindering the establishment of unified benchmarks. Moreover, the limited information content in pure molecular graphs or SMILES representations constrains their applicability to downstream tasks, particularly in structure-based drug design scenarios such as protein-ligand binding prediction and target-aware ligand generation (see Section \ref{section:3-3}), where 3D spatial information is crucial for understanding molecular interactions.

\subsection{Target-aware Generation}\label{section:3-3}
Target-aware molecule generation has emerged as one of the most active areas of DM research,\cite{anderson2003process} which focuses on generating molecules, specifically ligands, tailored to interact with a target protein to accelerate drug discovery, rather than exploring vast chemical spaces. Representing the 3D structure of the protein and generating valid, stable ligands that bind to a pocket of the protein are ongoing challenges in this field. 

Formally, the diffusion process is conditioned on the protein pocket as $p_\theta(x_t^{(L)}|x_{t-1}^{(L)}, x^{(P)})$ where $x^{(P)}$ represents the protein pocket and $x^{(L)}$ denotes the generated ligand that is encoded as atom positions and their atom types similar to EDM.\cite{hoogeboom2022equivariant} 
Two fundamental approaches for conditional DMs—--classifier-based guidance and classifier-free guidance, share a common theoretical basis. They incorporate conditional information through an auxiliary function $p_\phi$, modifying the diffusion process as $ p_\theta(x_t^{(L)}|x_{t-1}^{(L)}, x^{(P)}) \propto p_\theta(x_t^{(L)}|x_{t-1}^{(L)}) \cdot p_\phi(x^{(P)}|x_t^{(L)}) $. Here, $ p_\theta(x_t^{(L)}|x_{t-1}^{(L)})$ represents the unconditional diffusion process for generating ligands, while $p_\phi(x^{(P)}|x_t^{(L)})$ acts as a guidance signal that evaluates the compatibility between the generated ligand and the protein pocket. The diffusion process is a combination of these two components. A critical limitation of conditional DMs is that they treat protein pockets as static structures, failing to capture dynamic conformational changes during binding.

Key considerations include ensuring the chemical feasibility of generated molecules, effectively capturing the spatial interactions between the protein and ligand atoms, and striking a balance between exploring chemical space for molecule diversity and exploiting existing knowledge of protein-ligand interactions. The most widely used benchmark dataset for this purpose is the CrossDocked2020 dataset.\cite{francoeur2020three}  It contains 22.5 million ligands and their poses docked into protein pockets from the Protein Data Bank,\cite{pdb} and the subset showing high binding affinity is typically used to train models.

\paragraph{Basic Models}
Prior approaches, such as PocketMol\cite{gilder2001pocketmol} and GraphBP,\cite{liu2022generating} generate molecules autoregressively by adding one atom at a time, which are often used as baselines for comparison to DMs that are non-autoregressive. These autoregressive methods suffer from error accumulation and lack global molecular structure optimization.
TargetDiff\cite{guan20233d_targetdiff} generates ligands using an SE(3)-equivariant DM. It represents both ligand atoms and the protein pocket as a 3D point cloud where points are annotated with features such as atom types and chemical properties. The protein pocket information is encoded as conditional input to an EGNN\cite{satorras2021n} during the diffusion process.

Concurrently with TargetDiff, DiffSBDD\cite{schneuing2022structure} introduced an inpainting strategy for ligand generation, inspired by image inpainting techniques such as Repaint.\cite{lugmayr2022repaint} It first trains an unconditioned generative model on ligand-protein complexes and then generates ligands by fixing the protein pocket part of the complex. This strategy allows existing pre-trained models to be adapted for target-aware molecule generation. However, both methods represent primitive approaches that lack explicit binding affinity optimization and molecule-specific adaptations.

\paragraph{Conditioning Methods in Molecule Generation}
Conditioning Techniques attempt to build on existing models and incorporate advanced conditioning strategies to tailor molecular outputs to more specific requirements, such as improving binding affinity or preserving pharmacophore features. SILVR\cite{runcie2023silvr} utilizes existing target-free diffusion-based models such as EDM\cite{hoogeboom2022equivariant} by introducing a conditioning method that allows for the generation of molecules fitting specific protein binding sites without retraining the model. Inspired by ILVR\cite{choi2021ilvr} in image generation, SILVR mixes unconditionally generated samples with a reference molecule---typically one that fits the target protein well---in the latent space, offering control over the degree of conditioning. A notable limitation is the lack of control over specific conditions, restricts its ability to prioritize binding affinity improvement over merely structural similarity to the reference molecule. MolSnapper\cite{ziv2024molsnapper} extends the MolDiff\cite{peng2023moldiff} framework by integrating pharmacophore information (e.g., functional groups or spatial arrangements) to guide generation. By fixing pharmacophore positions and types as conditional inputs during generation, MolSnapper can generate molecules with high binding affinity with better preservation of molecular validity. ShapeMol,\cite{chen2023shape} building on TargetDiff, focuses on generating molecules with 3D shapes that closely matches a target molecule. This method encodes the target molecule's surface shape, represented as a polygon mesh to capture detailed geometric features, into latent embeddings. The embeddings serve as conditional inputs during the generation process, ensuring shape similarity. These methods illustrate a trend towards using geometric or functional conditions to refine molecule generation. They address the challenge of applying molecule generation to requirements that are difficult to pre-define, with the aim of improving the utility of future pre-trained models. However, these highly specialized conditioning approaches raise concerns about scalability and practical applicability, as they may be over-engineered for specific scenarios while lacking broader utility in real-world drug discovery pipelines.

\paragraph{Incorporating Binding Affinity in Generation}
This group of work aims to improve protein-ligand binding affinity by incorporating binding affinity predictors like AlphaFold3\cite{abramson2024accurate} into the DM. KGDiff\cite{qian2024kgdiff} takes a direct approach by creating an expert network to predict binding affinity and using the prediction to guide molecule generation.\cite{dhariwal2021diffusion} IPDiff\cite{huang2023protein} uses IPNet, a binding affinity predictor trained on noisy protein-ligand data. It introduces \textit{prior-shifting}, a mechanism that utilizes interaction representations from IPNet, to directly modify the diffusion process on atom coordinates. At each diffusion step, a shift is added to the atom coordinates, steering the generation toward higher-affinity molecules. PromptDiff\cite{yang2023prompt} employs a Protein-Molecule Interaction Network (PMINet) for two purposes: assessing the binding affinity of generated molecules and retrieving top-ranked candidates from a 3D molecule database based on their predicted affinity. These candidates are then used as conditional inputs to refine the subsequent diffusion steps. However, incorporating binding affinity introduces error propagation, as models may optimize for computational scores rather than realistic affinity. Reinforcement learning approaches\cite{ouyang2022training} show promise in substantially alleviating this issue.

\paragraph{Protein-Ligand Interaction Models.}
Several innovative approaches have been developed to better represent protein-ligand interactions. They utilize protein pocket information to guide ligand generation. Building on MDM,\cite{huang2023mdm} PMDM\cite{huang2024dual} extends the local and global encoders architecture  to condition on specific protein binding sites, enhancing the modeling of local and global interactions. Similarly,  InterDiff,\cite{huang2023protein} also building on MDM, improves the model-based approach by dynamically predicting interactions among atoms rather than using pre-defined distance thresholds. BindDM\cite{huang2024binding} identifies a key subcomplex within the protein-ligand complex, serving as a learned representation of protein-ligand interactions. The subcomplex is dynamically updated at each timestep based on the evolving full complex, enabling more precise forms for protein-ligand binding. AliDiff\cite{gu2024aligning} utilizes target proteins and desired interactions along with reward functions on user-defined objectives to define a reinforcement learning framework, extending the capabilities of IPDiff\cite{huang2023protein}. AliDiff introduces a theoretically grounded mechanism for aligning the generation process with precise energy-based optimization objectives. D3FG\cite{lin2024functional} represents molecules by decomposing them into rigid functional groups connected by mass-point linkers, which model the degrees of freedom of molecular structures. Using a pre-defined list of functional groups,  ligand-protein interactions can be better modeled by explicitly encoding chemical interactions. On the other hand, DecompDiff\cite{guan2024decompdiff} improves ligand-protein binding affinity by decomposing ligands into functional arms and a central scaffold. The arms interact with both the protein pocket and the scaffold, while the scaffold interacts exclusively with the arms. These interactions are modeled through a k-nearest neighbors (kNN) graph connecting ligand and protein atoms to capture spatial proximity.  While these sophisticated interaction modeling approaches show promise, they typically rely on rigid predefined interaction patterns that may not generalize well across diverse protein families and binding modes.

%use attachment points to replace anchor point
\paragraph{Fragment-based Models}
%Fragment-based methods 
Another set of methods combines autoregressive and nonautoregressive approaches by decomposing molecules into fragments: first selecting appropriate fragments, and then attaching them autoregressively. This two-step process focuses on two key operations---identifying attachment points on the molecular structure that is currently generated and creating suitable fragments in a DM for each position. Fragment-based methods incorporate pre-defined chemical bonds within their molecular fragments, allowing direct assembly of complete molecules without requiring additional bond formation steps. DiffLinker\cite{igashov2024equivariant} generates a single linker to connect multiple input fragments. The method addresses two key aspects: determining the attachment points of the input fragments and how to construct a linker that fulfills both functional requirements and spatial constraints. This approach adapts the EDM\cite{hoogeboom2022equivariant} framework  to focus specifically on fragment connection tasks.
LinkerNet\cite{guan2024linkernet} enhances DiffLinker by explicitly modeling the pose prediction of the connected fragments, as the rotational degrees of freedom impact linker quality. AutoFragDiff\cite{ghorbani2023autoregressive} applies DMs to generate fragments based on a provided scaffold and attachment points. At each step, it relies on a separate model to predict attachment points and samples fragments to be added at these points using DMs. During this process, Lennard-Jones interaction scores are used as classifier guidance\cite{ho2022classifier} to optimize the interactions between the pocket and the generated fragments. In contrast, AUTODIFF\cite{li2024autodiff} selects motifs, which represent specific molecular substructures, and uses DMs to generate the torsional angles of the added motifs. AUTODIFF also relies on separate models to predict attachment points and to select the motifs. DecompOPT\cite{zhou2024decompopt} extends the ligand decomposition approach of DecompDiff\cite{guan2024decompdiff} by introducing a multi-stage strategy. It first generates a ligand, decomposes it into fragments, and iteratively regenerates improved ligands using these fragments as a reference set to optimize binding affinity with the protein pocket.  While fragment-based models are appealing due to their human-like approach to molecular assembly and potentially higher interpretability, they suffer from complex multi-stage pipelines that are difficult to scale, heavy dependence on fragment definitions, and notably poor validity as shown in Table~\ref{table:combined_results}.

\begin{figure}[!ht]
    \centering
    \includegraphics[width=0.45\textwidth]{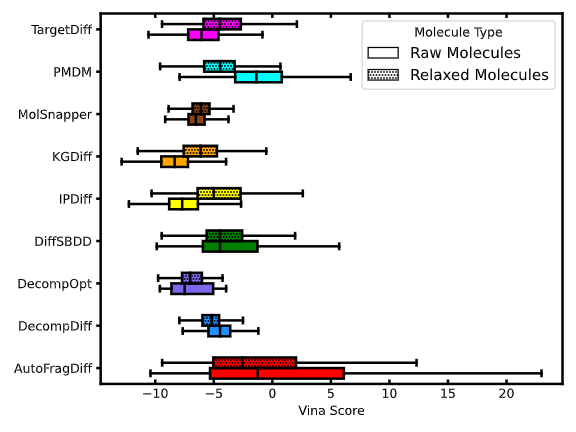}
    \caption{Comparison of Vina scores between raw and relaxed molecules. A decreased score indicates relaxation improves molecular validity (lower scores are better), while an increased score suggests the model overfits to Vina scores.}
    \label{vina}
\end{figure}
% \refstepcounter{figure}
% \label{fig:vina}

\begin{table*}[ht]
\centering
\setlength{\tabcolsep}{3pt}
\begin{subtable}[t]{\textwidth}
\centering
\begin{tabular}{|c|c|c|c|c||c|c|c|}
\hline
\textbf{dataset} & \multicolumn{4}{c|}{\textbf{QM9}} & \multicolumn{3}{c|}{\textbf{GEOM-Drugs}} \\
\hline
\textbf{Model} & \textbf{MS} $\uparrow$ & \textbf{Validity} $\uparrow$& \textbf{Uniqueness $\uparrow$} & \textbf{Novelty} $\uparrow$ & \textbf{AS} $\uparrow$ & \textbf{Validity} $\uparrow$ & \textbf{MS} $\uparrow$ \\
\hline
EDM\cite{hoogeboom2022equivariant} & 78.7 & 89.2 & 98.6 & 67.4 & 81.3 & 83.5 & 0 \\
GeoLDM\cite{xu2023geometric} & 88.9 & 94.2 & 98.0 & 59.5 & 84.4 & 99.3 & 0 \\
VoxMol\cite{o20243d_voxmol} & 90.1 & 98.6 & 98.9 & 100 & 99.0 & 94.7 & - \\
FMG\cite{dumitrescu2024field} & 91.5 & 98.8 & 68.8 & - & 94.7 & 64.9 & - \\
HierDiff\cite{qiang2023coarse} & 100 & 87.8 & 97.9 & 83.6 & - & 94.0 & - \\
% HierDiff$_P^*$ & \textbf{100} & 90.4 & 99.5 & 83.2 & 100 & 90.4 & - \\
MiDi\tnote{2}\cite{vignac2023midi} & 97.5 & 97.9 & 97.6 & 67.5 & 99.8 & 77.8 & 91.6 \\
MDM\cite{huang2023mdm}& 91.9 & 98.6 & 94.6 & 90.0 & 99.0 & 99.5 & 62.2 \\
MolDiff\cite{peng2023moldiff} & - & 97.0 & 99.1 & 100.0 & - & 94.7 & - \\
GFMDiff\cite{xu2024geometric} & 87.7 & 96.3 & 95.1 & - & 86.5 & - & - \\
\hline
\end{tabular}
% \caption{Target-free Molecular Generation Results}
\label{table:sub-a}
\end{subtable}
% \hfill
\begin{subtable}[t]{\textwidth}
\centering
\begin{tabular}{|l|cc|c|c|cc|cc|}
\hline
\multirow{2}{*}{\textbf{Model}} & \multicolumn{2}{c|}{\textbf{Validity3D (\%)} $\uparrow$} & \multirow{2}{*}{\textbf{QED} $\uparrow$} & \multirow{2}{*}{\textbf{SA Score} $\uparrow$} & \multicolumn{2}{c|}{\textbf{Vina} $\downarrow$} & \multicolumn{2}{c|}{\textbf{Strain Energy} $\downarrow$} \\
\cline{2-3}
\cline{6-9}
 & \textbf{Normal} & \textbf{Relaxed} &  &  & \textbf{Normal} & \textbf{Relaxed} & \textbf{Normal} & \textbf{Relaxed} \\
\hline
DecompDiff\cite{guan2024decompdiff}          & 1.1\% & 47.0\% & 0.432 & 4.541  & -3.919  & -4.337 & 385.4  & 32.6 \\
MolSnapper\cite{ziv2024molsnapper}           & 0.1\% & 37.1\% & 0.326 & 4.223  & 2.925 & -2.728 & 321.0   & 41.0 \\
PMDM\cite{huang2024dual}                     & 6.2\% & 2.9\%  & 0.628 & 7.285  & -2.830   & -5.042 & 172.1  & 19.3  \\
TargetDiff\cite{guan20233d_targetdiff}       & 0.9\% & 42.0\% & 0.443 & 4.816   & -5.633  & -4.935 & 369.5  & 33.8 \\
IPDiff\cite{huang2023protein}                & 0.2\% & 40.2\% & 0.513 & 5.157  & -6.611  & -5.357 & 958.5    & 34.7 \\
DiffSBDD\cite{schneuing2022structure}        & 1.2\% & 44.8\% & 0.485 & 4.514  & -4.377  & -4.324 & 355.7    & 25.7 \\
KGDiff\cite{qian2024kgdiff}                  & 0.7\% & 42.9\% & 0.521  & 5.087 & -7.021  & -5.557 & 3812.1 & 38.1 \\
AutoFragDiff\cite{ghorbani2023autoregressive}  & 2.4\% & 40.2\% & 0.437 & 4.431  & 50.506  & 49.145 & 380.9  & 44.8 \\
DecompOPT\cite{zhou2024decompopt}            & 0.0\% & 36.2\% & 0.266   & 5.004    & -6.016   & -4.834  & 518.6   & 42.4\\
\hline
\end{tabular}
% \caption{Target-aware Molecular Generation Results}
\label{table:sub-b}
\end{subtable}
\caption{Experimental evaluation of DMs for target-Free and target-aware molecule generation. 
(Top) Target-free molecule generation metrics on the QM9\cite{ramakrishnan2014quantum} and GEOM-Drugs.\cite{axelrod2022geom} datasets %Works with * means the result is reported. 
(Bottom) Target-aware molecule generation metrics using Genbench3D,\cite{baillif_genbench3d} including Validity3D (Normal and Relaxed), QED, SA Score, Vina Score (Normal and Relaxed), and Strain Energies (Normal and Relaxed). Each model sampled 1000 molecules for five proteins, and the metrics are evaluated on these molecules.}% Relaxation is performed in order to increase the number of valid molecules.}
\label{table:combined_results}
\end{table*}

\refstepcounter{table}
\label{table:combined_results}
\section{Benchmarking Representative DMs}
\label{section:benchmark}
We conducted extensive experiments to evaluate eighteen representative DMs, nine for target-free generation and nine  for target-aware generation across three benchmark datasets. For target-free generation, we used the QM9\cite{ramakrishnan2014quantum} and GEOM-Drugs\cite{axelrod2022geom} datasets, while for target-aware generation, the CrossDocked2020\cite{francoeur2020three} is the benchmark dataset. 

For target-free generation, we employed five widely adopted  evaluation metrics: Validity, Uniqueness, Atom Stability (AS), Molecule Stability (MS), and Novelty. These metrics were implemented using RDKit\cite{rdkit} and follow the definitions established in prior work.\cite{garcia2021n} From Table \ref{table:combined_results}, we observe MiDi has the best performance, with a high Stability, Validity and competitive Novelty. Notably, MiDi’s generation of 3D molecular graphs contributes to its superior results. Additionally, we note that voxel-based approaches such as VoxMol and FMG require significantly more training time compared to other methods. This trade-off highlights the importance of innovative methods like MiDi, which achieve state-of-the-art results with practical training efficiency.

Target-aware generation is the focus of our benchmarking, and a latest benchmarking approach, Genbench3D,\cite{baillif_genbench3d} is used. Validity3D,\cite{baillif_genbench3d} Vina Score\cite{eberhardt2021autodock}, Quantitative Estimate of Druglikeness (QED),\cite{bickerton2012quantifying} Synthetic  Accessibility (SA) Score,\cite{ertl2009estimation} and Strain Energy are used as evaluation metrics. Validity3D, introduced by Genbench3D, measures the percentage of three-dimensionally valid molecules by assessing key structural parameters (e.g., bond lengths, torsion angles, ring flatness) using the Cambridge Structural Database (CSD) as a reference. It computes the likelihood that these parameters fall within chemically reasonable ranges while accounting for potential steric clashes. The Vina Score, proposed by AutoDock Vina,\cite{eberhardt2021autodock} is a metric to measure binding affinity. The definitions of other metrics follow the established standards in Pocket2Mol.\cite{gilder2001pocketmol} Except for Validity3D, which is reported as a percentage, all other metrics are presented as averaged medians (e.g., the median binding affinity value of all generated molecules for a protein is calculated, and then the average of the median values across all proteins is reported).

We randomly selected five protein targets  (2Z3H, 2PQW, 5L1V, 2E24, and 4PXZ) from the CrossDocked test set and generated 1000 molecules for each protein target using each model. For each pool of 1000 sampled molecules from a given model and for a given protein, we conducted the benchmark comparison twice, once with and once without local relaxation. The relaxation process utilized RDKit's Merck Molecular Force Field (MMFF)  minimization to optimize ligand conformations. 

As shown in Table \ref{table:combined_results}, relaxation significantly improves Validity3D for most models, with the mean value increasing from 1.42\% to 37.0\%. PMDM is the only exception showing a unique decrease in geometric validity after relaxation. Comprehensive evaluations demonstrate varied responses to relaxation across different models. Notably, fragment-based approaches including AutoFragDiff and DecompOPT, performed poorly across all metrics. Based on our evaluation metrics for the post-relaxation results, KGDiff emerged as the best target-aware model. Furthermore, it is helpful to consider the estimated binding affinity as an additional validation of generated molecules, so we compare Vina Scores. 

Figure \ref{fig:vina} reveals distinct patterns in Vina Score before and after relaxation across the comparison  models. We observe that TargetDiff, KGDiff, and IPDiff showed performance deterioration after relaxation; DecompDiff, MolSnapper, DiffSBDD, AutoFragDiff, and DecompOPT maintained comparable performance levels; and PMDM alone demonstrated improved scores after relaxation. Performing relaxation on generated molecules highlights a trade-off between Vina Score and Validity3D under the current approaches. Although relaxation might produce geometries that are more likely to be valid, the ability of the processed molecules to bind to their targets is often weakened. Finally, PMDM's unique performance improvement in Vina Score after relaxation indicates that model architecture design directly affects the ability of generated structures to maintain their properties during force field optimization.

%PMDM performed best on Validity3D before relaxation, while DecompDiff excelled after relaxation. KGDiff consistently scored highest on Vina score, and PMDM demonstrated the best Strain Energy performance.

\begin{figure}
    \centering
    \includegraphics[width=\columnwidth]{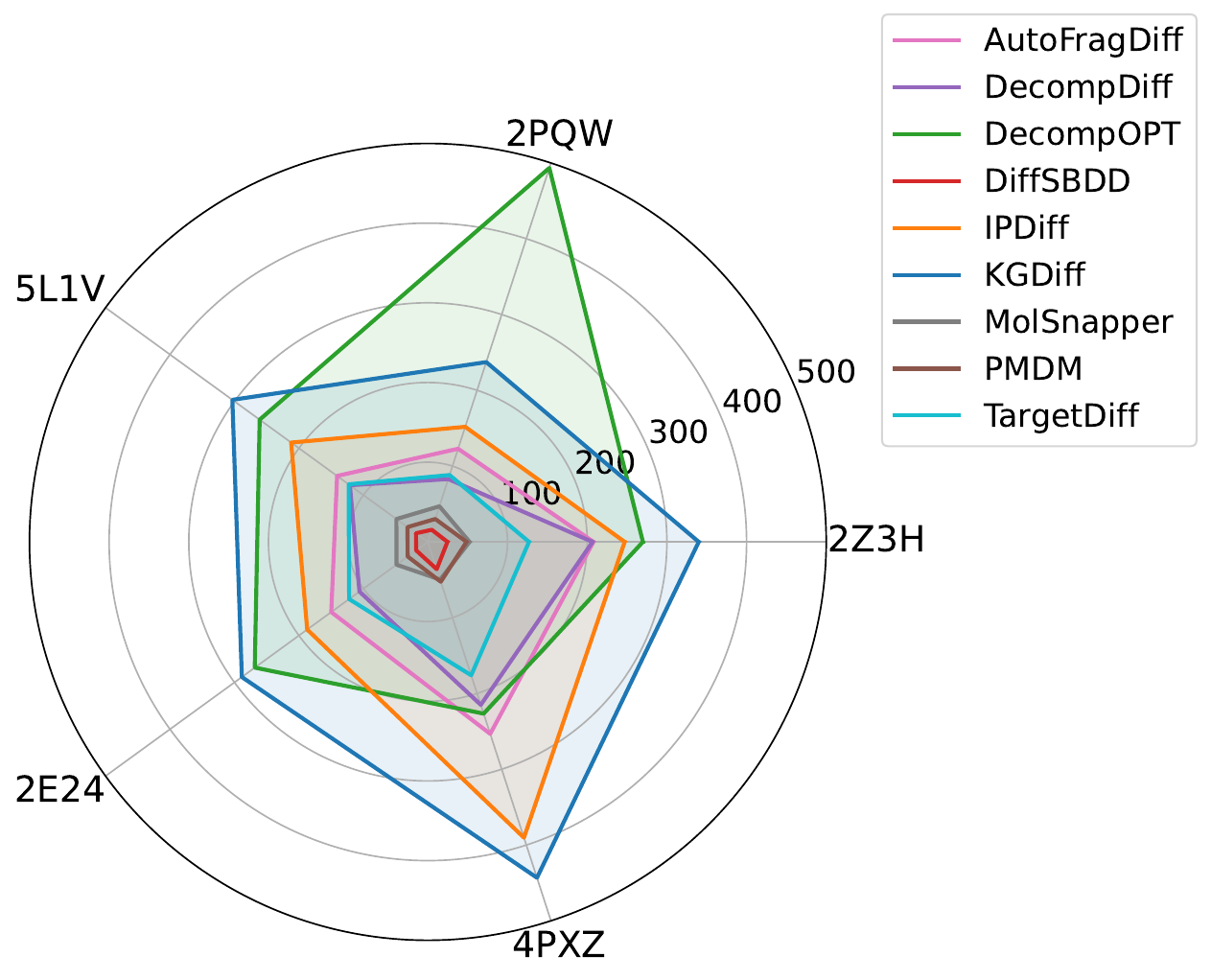}
    \caption{Star plot showing the averaged execution time (in seconds) for a batch of 10 molecules across five different proteins. For each protein and each model, we generate 10 batches of 10 molecules and take the average execution time.}
    \label{fig:starplot}
\end{figure}

The execution times of each of the models we tested can be seen in Figure \ref{fig:starplot}. For each model and each protein, we sampled 10 batches of 10 molecules and plotted the average execution time.  We observe very little variance in the execution across the batches for any given model-protein pair. We note that DecompOPT was the slowest for 2PQW while the rest of the models were slowest on 4PXZ. For 2PQW, DecompOPT generated a ligand with significantly more atoms than for the other proteins.

\section{Limitations}
% DMs have demonstrated significant potential from de novo generation to target-based design. However, DMs' black-box nature can hinder interpretability, making it difficult to understand molecule generation mechanisms and potential biases.
%Although DMs have achieved remarkable progress in generating complex molecular structures, like most deep learning-based methods, they have a significant limitation: their inherent black-box nature. This characteristic presents significant challenges for model improvement and for correcting potential biases in the generated molecules. For example, in 3D molecule generation, it is difficult to understand how the model determines specific atomic coordinates and bond angles during the reverse diffusion process. % Moreover, the generated molecules often tend to be biased towards certain types in the training set, restricting the diversity of the chemical space and potentially missing novel drug candidates.
Current molecular DMs suffer from method-specific interpretability challenges that hinder systematic improvement. Fragment-based approaches lack transparency in fragment selection logic, while coordinate-based models provide no insight into how geometric constraints are learned or enforced during generation. Graph-based methods, despite strong performance, offer limited understanding of chemical rule acquisition during node-edge operations. The scientific community expects future DMs to integrate explainable AI to clarify how and why specific molecules are generated—such as by visualizing diffusion steps or highlighting key atom and bond decisions. Enhanced interpretability fosters trust and validation, which is vital for advancing their use in drug discovery.%The scientific community expects future DMs to incorporate an explainable AI component to disclose how and why certain molecules are generated. This could involve visualizing the diffusion steps and presenting key decisions related to atom and bond addition or removal. Improving model interpretability helps researchers better understand and trust the generated molecules, which is crucial for validating and promoting their applications, especially in drug discovery.

% The current evaluation landscape for molecular generation remains vulnerable. {\bf do you mean benchmark datasets not benchmark approach?} Despite the introduction of advanced benchmarks like PoseButter\cite{buttenschoen2024posebusters} and PoseCheck\cite{harris2023benchmarking}, their adoption remains limited. Traditional benchmark datasets such as QM9 and CrossDocked2020, along with conventional metrics such as QED, SA Score, Vina Score, and other RDkit-based metrics, continue to dominate the field for validating the generation. However, these approaches offer limited insights, failing to comprehensively capture the multifaceted nature of drug discovery, such as pharmacokinetic properties (e.g., absorption, distribution, metabolism, and excretion), toxicity profiles, physicochemical properties.
% Crucially, there is distinct lack of connection between computational drug design and real-world drug development criteria. The correlation between current computational metrics and critical factors such as clinical success rates or therapeutic efficacy has not been systematically investigated. This gap highlights the urgent need for more sophisticated evaluation frameworks that can bridge theoretical molecular generation with the practical complexities of pharmaceutical research. {\bf this is only about evaluation, and it is true for all deep models, so what is specific limitation of DM?}
Furthermore, the gap between theoretical molecule generation and the practical application of pharmaceutical research demands more sophisticated evaluation frameworks. The reliance on post-hoc geometric relaxation to achieve structural validity reveals that training objectives fail to incorporate sufficient chemical constraints. In some cases, a generated molecule appears geometrically valid but has unfavorable energetic or conformational properties, making it biologically ineffective. Overall, evaluating DMs for small molecule generation requires a comprehensive approach that integrates theoretical and practical aspects to drive progress.

Computational efficiency varies significantly across molecular DM architectures. While unconditional molecular DMs achieve competitive speeds, target-aware approaches like KGDiff require substantially longer processing times (30.3 seconds vs. 2.24 seconds for MolT5),\cite{qian2024kgdiff, edwards2022translation} representing a 15-fold computational overhead compared to text-based models due to protein structure encoding and interaction modeling. Promising acceleration strategies include acceleration sampling methods for DMs, adaptive sampling based on molecular complexity, and knowledge distillation to more efficient architectures.\cite{lipman2022flow, song2023consistency, sauer2024fast}

\section{Future Perspectives}

Building on these limitations, future research should focus on enhancing the physical realism and biological relevance of molecular diffusion models. Integrating molecular mechanics (MM) or quantum chemistry (QC) energy constraints directly into diffusion processes offers a promising path toward generating conformations that are both structurally valid and energetically favorable, reducing the current reliance on post-hoc relaxation.

The increasing availability of high-quality protein structure predictions, exemplified by AlphaFold3,\cite{abramson2024accurate} creates new opportunities for advancing multi-modal and target-aware molecular generation. Joint modeling of molecules and their biological targets, combined with improved evaluation frameworks that incorporate synthetic accessibility, binding relevance, and experimental feasibility, will better align model outcomes with real-world drug discovery needs.

AI-driven drug discovery is already demonstrating early clinical impact. For example, INS018\_055,\cite{ren2025small} an AI-designed candidate for idiopathic pulmonary fibrosis, advanced to Phase II trials within 18 months. Looking ahead, the ability of generative models to augment or partially replace traditional experimental workflows will be critical for achieving broader impact. Continued progress will depend on both methodological advances and the availability of high-quality structural and biochemical data to support reliable model training and validation.

\section{Conclusion}
% Generative AI is promising to revolutionize the drug discovery workflow. Small-molecule generation, in particular, demands profound chemical knowledge. Existing DMs still struggle with generalizability when creating valid and synthesizable molecular structures. 
Generative AI, particularly diffusion models (DMs), is transforming drug design for molecular geometry and protein pocket–aware molecule generation. This review highlights both breakthroughs and bottlenecks in current molecular DMs, including data sparsity, lack of physics integration, unreliable evaluation metrics, and scalability limitations. Current models often produce energetically invalid conformations or misaligned binding poses; in our evaluation, MiDi\cite{vignac2023midi} demonstrated superior performance in stability and validity for unconditional generation, while KGDiff\cite{qian2024kgdiff} and PMDM\cite{huang2024dual} achieved the best target-aware performance, yet all models still require geometric correction, underscoring fundamental gaps in current 3D generation capabilities. Looking ahead, advancing physics-informed DMs that incorporate differentiable force fields or hierarchical learning frameworks will be important for generating physically realistic and biologically meaningful molecules. Cross-disciplinary collaboration will be essential to further translate these computational advances into practical drug discovery applications.

%Recently, the release of AlphaFold3,\cite{abramson2024accurate} along with its support for protein-ligand docking functionality, has paved the way for more precise and efficient evaluation of generated small molecules. We hope to see models that conform to the laws of structural chemistry and quantum mechanics. The rapidly increasing availability of data can contribute to the translation of computational predictions into physiological progress.

\section{Declarations of interest} 
Auhtors have no conflict of interests to declare.

\section{Acknowledgments}
This work was partially supported by an endowment fund from the Frederick H Leonhardt Foundation Inc. to JB for improving impacts of computer science techniques on real-world applications.
% %\vskip3pt
\appendix
\printcredits
% \bibliographystyle{model1-num-names}  % Elsevier 
\bibliographystyle{cas-model2-names}  % 
% \bibliographystyle{numcompress}

\bibliography{cas-refs}

\end{document}